\documentclass[%
 preprint,
 amsmath,amssymb,
 aps,
 superscriptaddress,
12pt]{revtex4-1}
\usepackage{lmodern}
\PassOptionsToPackage{hyphens}{url}
\usepackage{graphicx}
\usepackage{dcolumn}
\usepackage{bm}
\usepackage[utf8]{inputenc}
\usepackage{amssymb}
\usepackage{amsmath}
\usepackage{bm}
\usepackage[dvipsnames]{xcolor}
\usepackage{placeins}
\usepackage{listings}
\usepackage{setspace}
\usepackage[normalem]{ulem}
\usepackage{xr}

\usepackage{array}
\usepackage{url}
\usepackage{breakurl}

\newcolumntype{P}[1]{>{\centering\arraybackslash}p{#1}}
\newcolumntype{M}[1]{>{\centering\arraybackslash}m{#1}}

\graphicspath{{./figures/}}

\begin{document}

\title{Risk mapping for COVID-19 outbreaks in Australia using mobility data}

\author{Cameron Zachreson}
\affiliation{School of Computing and Information Systems, The University of Melbourne, Australia}
\author{Lewis Mitchell}
\affiliation{School of Mathematical Sciences, The University of Adelaide, Australia}
\author{Michael J. Lydeamore}
\affiliation{Victorian Department of Health and Human Services, Government
of Victoria, Australia}
\affiliation{Department of Infectious Diseases, The Alfred and Central Clinical School, Monash University, Melbourne, Australia}
\author{Nicolas Rebuli}
\affiliation{School of Public Health and Community Medicine, University of New South Wales, Australia}
\author{Martin Tomko}
\affiliation{Melbourne School of Engineering, The University of Melbourne, Australia}
\author{Nicholas Geard}
\affiliation{School of Computing and Information Systems, The University of Melbourne, Australia}
\affiliation{The Peter Doherty Institute for Infection and Immunity, The Royal Melbourne Hospital and The University of Melbourne, Australia}

\begin{abstract}

COVID-19 is highly transmissible and containing outbreaks requires a rapid and effective response. Because infection may be spread by people who are pre-symptomatic or asymptomatic, substantial undetected transmission is likely to occur before clinical cases are diagnosed. Thus, when outbreaks occur there is a need to anticipate which populations and locations are at heightened risk of exposure. In this work, we evaluate the utility of aggregate human mobility data for estimating the geographic distribution of transmission risk. We present a simple procedure for producing spatial transmission risk assessments from near-real-time population mobility data. We validate our estimates against three well-documented COVID-19 outbreaks in Australia. Two of these were well-defined transmission clusters and one was a community transmission scenario. Our results indicate that mobility data can be a good predictor of geographic patterns of exposure risk from transmission centres, particularly in outbreaks involving workplaces or other environments associated with habitual travel patterns. For community transmission scenarios, our results demonstrate that mobility data adds the most value to risk predictions when case counts are low and spatially clustered. Our method could assist health systems in the allocation of testing resources, and potentially guide the implementation of geographically-targeted restrictions on movement and social interaction. 

\end{abstract}

\maketitle

\section{Introduction}

Similar to other respiratory pathogens such as influenza, the transmission of {SARS-CoV-2} occurs when infected and susceptible individuals are co-located and have physical contact, or exchange bioaerosols or droplets \cite{kutter2018transmission,siegel20072007}. Behavioural modification in response to symptom onset (i.e., self-isolation) can act as a spontaneous negative feedback on transmission potential by reducing the rate of such contacts, making epidemics much easier to control and monitor. However, COVID-19 (the disease caused by SARS-CoV-2 virus) has been associated with relatively long periods of pre-symptomatic viral shedding (approximately 5 - 10 days), during which time case ascertainment and behavioural modification are unlikely \cite{lauer2020incubation,he2020temporal}. In addition, many cases are characterised by mild symptoms, despite long periods of viral shedding \cite{young2020epidemiologic}. Transmission studies have demonstrated that asymptomatic and pre-symptomatic transmission hamper control of SARS-CoV-2 \cite{ferretti2020quantifying,li2020substantial,wei2020presymptomatic}. Pre-symptomatic and asymptomatic transmission has also been documented systematically in several residential care facilities in which surveillance was essentially complete \cite{arons2020presymptomatic,kimball2020asymptomatic}. Currently, there are no prophylactic pharmaceutical interventions that are effective against SARS-CoV-2 transmission. Therefore, interventions based on social distancing and infection control practices have constituted the operative framework, applied in innumerable variations around the world, to combat the COVID-19 pandemic.  

Social distancing policies directly target human mobility. Therefore, it is logical to suggest that data describing aggregate travel patterns would be useful in quantifying the complex effects of policy announcements and decisions \cite{buckee2020aggregated}. The ubiquity of mobile phones and public availability of aggregated near-real-time movement patterns has led to several such studies in the context of the ongoing COVID-19 pandemic \cite{pepe2020covid,martin2020effectiveness,bourassa2020social}. One source of mobility data is the social media platform Facebook, which offers users a mobile app that includes location services at the user's discretion. These services document the GPS locations of users, which are aggregated as origin-destination matrices and released for research purposes through the Facebook Data For Good program. The raw data is stored on a temporary basis and aggregated in such a way as to protect the privacy of individual users \cite{maas2019facebook}. Several studies have utilised subsets of this data for analysis of the effects of COVID-19 social distancing restrictions \cite{bonaccorsi2020economic,lee2020job,holtz2020interdependence,galeazzi2020human}. In addition, other sources of mobility data have been used to quantify the positive association between human travel, case importation, and local prevalence of COVID-19 after the disease emerged in the Chinese city of Wuhan and subsequently spread to other regions \cite{kraemer2020effect,jia2020population,niehus2020using}. 

In this work, we complement these studies by addressing the question: to what degree can real-time mobility patterns estimated from aggregate mobile phone data inform short-term predictions of COVID-19 transmission risk? Here, we examine outbreaks and population flows approximately three orders of magnitude smaller than those investigated previously in studies focused on the Chinese context \cite{kraemer2020effect,jia2020population}. On this scale, it is less clear whether strong associations between mobility and transmission risk will still be observable from the available data.

To address this question, we develop a straight-forward procedure to generate a relative estimate of the spatial distribution of future transmission risk based on current case data or locations of known transmission centres. To critically evaluate the performance of our procedure, we retrospectively generate risk estimates based on data from three outbreaks that occurred in Australia when there was little background transmission. We do not attempt to compute precise forecasts or predictions of case incidence, which would require a transmission model. Instead, we focus on differences in observed case counts between regions and investigate the degree to which they correlate with differences in observed mobility patterns. Our intention is to examine the utility of aggregate mobility data in generating spatial assessments of outbreak risk without precise definitions or models of disease dynamics. While we acknowledge that first-order factors determining transmission between infected and susceptible hosts may dominate local disease dynamics, the hypothesis motivating this work is that mobility between locations is an important determinant with respect to transmission over large, spatially distributed populations.

The initial wave of infections in Australia began in early March, 2020, and peaked on March 28th with 469 new cases. The epidemic was suppressed through widespread social distancing measures which escalated from bans on gatherings of more than 500 people (imposed on March 16th) to a nation-wide ``lockdown" which began on March 29th and imposed a ban on gatherings of more than 3 people. By late April, daily incidence numbers had dropped to fewer than 10 per day \cite{Aus_COVID}. The outbreaks we examine occurred during the subsequent period over which these general suppression measures were progressively relaxed. One of these occurred in a workplace over several weeks, one began during a gathering at a social venue, and one was a community transmission scenario with no single identified outbreak center, which marked the beginning of Australia's ``second wave" (which is ongoing as of August, 2020). The term ``community transmission" refers to situations in which multiple transmission chains have been detected with no known links identified from contact tracing and no specific transmission centres are clearly identifiable. 


In each case, we use the Facebook mobility data that was available during the early stages of the outbreak to estimate future spatial patterns of relative transmission risk. We then examine the degree to which these estimates correlate with the subsequently observed case data in those regions. Our results indicate that the accuracy of our estimates varies with outbreak context, with higher correlation for the outbreak centred on a workplace, and lower correlation for the outbreak centred on a social gathering. In the community transmission scenario without a well-defined transmission locus, we compare the risk prediction based on mobility data to a null prediction based only on active case numbers. Our results indicate that mobility is more informative during the initial phases of the outbreak, when detected cases are spatially localised and many areas have no available case data.

\section{Methods}

Our general method is to use an Origin-Destination (OD) matrix based on Facebook mobility data to estimate the diffusion of transmission risk based on one or more identified outbreak sources. The data provided by Facebook comprises the number of individuals moving between locations occupied in subsequent 8-hr intervals. For an individual user, the location occupied is defined as the most frequently-visited location during the 8-hr interval. More details on the raw data, the aggregation and pre-processing performed by Facebook before release, and our pre-processing steps can be found in the Supplementary Information.

COVID-19 case data is made publicly available by most Australian state health authorities on the scale of Local Government Areas (LGAs). In these urban and suburban regions, LGA population densities typically vary from approximately $0.2\times10^3$ to $5\times10^3$ residents per $\text{km}^2$, but can be low as $20$ residents per $\text{km}^2$ in the suburban fringe where LGAs contain substantial parkland and agricultural zones. The output of our method is a relative risk estimate for each LGA based on their potential for local transmission. The general method is as follows: 

\begin{enumerate}
\item{Construct the \emph{prevalence vector} $\bf{p}$, a column vector with one element for each location with a value corresponding to the transmission centre status of that location. For point-outbreaks in areas with no background transmission, we use a vector with a value of 1 for the location containing the transmission centre and 0 for all other locations. For outbreaks with transmission in multiple locations, we construct $\bf{p}$ using the number of active cases as reported by the relevant public health agency.}

\item{Construct an OD matrix $\bf{M}$, where the value of a component $M_{ij}$ gives the number of travellers starting their journey at location $i$ (row index) and ending their journey at location $j$ (column index). To approximately match the pre-symptomatic period of COVID-19, we average the OD matrix over the mobility data provided by Facebook during the 7 day period preceding the identification of the targeted transmission centre. By averaging over an appropriate time interval, the OD matrix is built to represent mobility during the initial stages of the outbreak, when undocumented transmission may have been occurring. The choice of appropriate time interval varied by scenario, as described below.}

\item{Multiply the OD matrix by the prevalence vector to produce an unscaled risk vector $\bf{r}$ with a value for each location corresponding to the aggregate strength of its outgoing connections to transmission centres, weighted by the prevalence in each transmission centre. This is re-scaled to give the relative transmission risk for each region $R_i$. In other words, we treat the OD matrix as analogous to the stochastic transition matrix in a discrete-time Markov chain, and compute the unscaled vector of risk values $\bf{r}$ as: 
\begin{equation}\label{eq_multi_risk}
    \bf{r} = \bf{M}\bf{p} \,, 
\end{equation}
so that $\bf{r}$ is approximately proportional to the average interaction rate between susceptible individuals from location $i$ and infected individuals located in the outbreak centres. These approximate interaction rates are then re-scaled to give relative risk values $R_i$ between 0 and 1: 
\begin{equation}\label{eq_multi_risk_2}
    R_i = \frac{r_{i}}{\sum_j {r_{j}}}\,.
\end{equation}

For point-outbreaks, this is simply:
\begin{equation}\label{eq_point_risk}
R_i = \frac{M_{ik}}{\sum_j M_{jk}} \,,
\end{equation}
where $k$ is the column index of the single outbreak location. The numerator is the number of individuals travelling from region $i$ to the outbreak centre, and the denominator is the total number of travellers into the outbreak centre over all origin locations $j$. 

In addition to the typical assumptions about equilibrium mixing (in the absence of more detailed interaction data), this interpretation is subject to the assumption that the strength of transmission in each centre is proportional to the number of active cases in that location. This assumption is consistent with the observation that the majority of individuals start and end their journeys in the same locations, but there is not sufficient data to unequivocally determine the relationship between transmission risk within an area and active case numbers in the resident population of that area. Therefore, it is appropriate to think of our method as a heuristic approach to estimating transmission risk based only on qualitative information about epidemiological factors and informed by near-real-time estimates of mobility patterns. These are derived from a biased sample of the population (a subset of Facebook users), and aggregated to represent movement between regions containing on the order of $10^3$ to $10^5$ individuals. 
}
\end{enumerate}

\subsection{Context-specific Factors}

Outbreaks occur in different contexts, some of which may suggest use of external data sources to infer at-risk sub-populations. Such inference can be used to refine spatial risk prediction. 

For example, the workplace outbreak we investigated occurred in a meat processing facility, where the virus spread among workers at the plant and their contacts. To adapt the general method to this context, we averaged OD matrices over the subset of our data capturing the transition between nighttime and daytime locations, as an estimate of work-related travel. In addition, we examined the effect of including industry of employment statistics as an additional risk factor. In this case, we used data collected by the Australian Bureau of Statistics (ABS) to estimate the proportion of meat workers by residence in each LGA, and weighted the outgoing traveller numbers by the proportion associated with the place of origin. 

The resulting relative risk value $R_i$ is a crude estimate of the probability that an individual:

\begin{itemize} 
\item{travelled from origin location $i$ into the region containing the outbreak centre;}
\item{travelled during the period when many cases were pre-symptomatic and no targeted intervention measures had been applied;}
\item{made their trip(s) during the time of day associated with travel to work and;}
\item{were part of the specific subgroup associated with the outbreak centre (in this case, those employed in meat-processing occupations).}
\end{itemize}

The variation described above is specific for workplace outbreaks in which employees are infected, but could be generally applied to any context where a defined subgroup of the population is more likely to be associated (e.g., school children, aged-care workers, etc.), or in which habitual travel patterns associated with particular times of day are applicable. In principle, this approach could be used to incorporate the effects of localised intervention policies or risk factors not directly related to mobility, such as limitations on gathering size, demographic factors affecting transmission risk (i.e., age distribution), or vaccination status of subpopulations. Here, in the context of the Cedar Meatworks outbreak, we focus on an occupation-related risk factor because of its assumed relationship with mobility between home and work. In the other two scenarios we investigate, no context-specific factors are incorporated.

While we make no explicit assumptions about spatial heterogeneity of disease dynamics, our choice to integrate mobility data for the 7 days preceding each risk estimate, along with our decision to validate these estimates against raw case reports implicitly assumes spatially homogeneous temporal lags between the events associated with transmission and those corresponding to subsequent case ascertainment. To test the potential sensitivity of our results to this implicit assumption, we examined the temporal autocorrelation of the mobility matrices used in our study (Figure \ref{OD_autocorr}). This analysis revealed a very high level of temporal consistency in relative mobility volumes between origin-destination pairs. Because mobility patterns are consistent in time, the risk estimate at time $t$ is not sensitive to the particular choice of integration interval as long as that interval is at least 7 days to account for weekly fluctuations in behaviour between weekend and weekday travel.

\section{Results}


For each of the three outbreak scenarios, we present the mobility-based estimates of the relative transmission risk distribution, and a time-varying correlation between our estimate and the case numbers ascertained through contact tracing and testing programs. For details of these correlation computations, see the Supplementary Information. 

\subsection{Cedar Meats}

\subsubsection{Scenario}


Cedar Meats is an abattoir (slaughterhouse and meat packing facility) in Brimbank, Victoria. It is located in the western area of Melbourne. It was the locus of one of the first sizeable outbreaks in Australia after the initial wave of infections had been suppressed through wide-spread physical distancing interventions. Meat processing facilities are particularly high-risk work environments for transmission of SARS-CoV-2, so it is perhaps unsurprising that the first large outbreak occurred in this environment \cite{dyal2020covid,richmond2020interregional}. It began at a time when community transmission in the region was otherwise undetected. As the transmission cluster grew, it was thoroughly traced and subsequently controlled. The contact-tracing effort included (but was not limited to) intensive testing of staff, each of which required a negative test before returning to work, 14-day isolation periods for all exposed individuals, and daily follow-up calls with every close contact. The outbreak was officially recognised on April 29th, when four cases were confirmed in workers at the site and, according to media reports, Victoria DHHS informed the meatworks of these findings \cite{BeefCentral}. The outbreak was first mentioned in the daily COVID-19 updates from Victoria DHHS on May 2nd, when the number of confirmed cases associated with the cluster had risen to eight \cite{DHHS_May2}. 

The Cedar Meats outbreak began when it was introduced into the workplace, where it subsequently spread to a large number of staff, and members of their households. We therefore selected for the distribution of travellers that may have been travelling \emph{to work} in the area of Cedar Meats during the period over which undetected transmission was likely. Specifically, we generated mobility risk maps based on trips {\textit{into}} the Brimbank region for the nighttime~$\rightarrow$~daytime OD matrix, averaged over the period between April 21st, and April 27th, 2020. We note that while there were only two SARS-CoV-2 positive cases associated with the cluster during this period (in two different areas), 43 cases were detected in the following week with infected individuals residing in 14 different locations. 

\begin{figure}
\centering
\includegraphics[width=0.7\columnwidth]{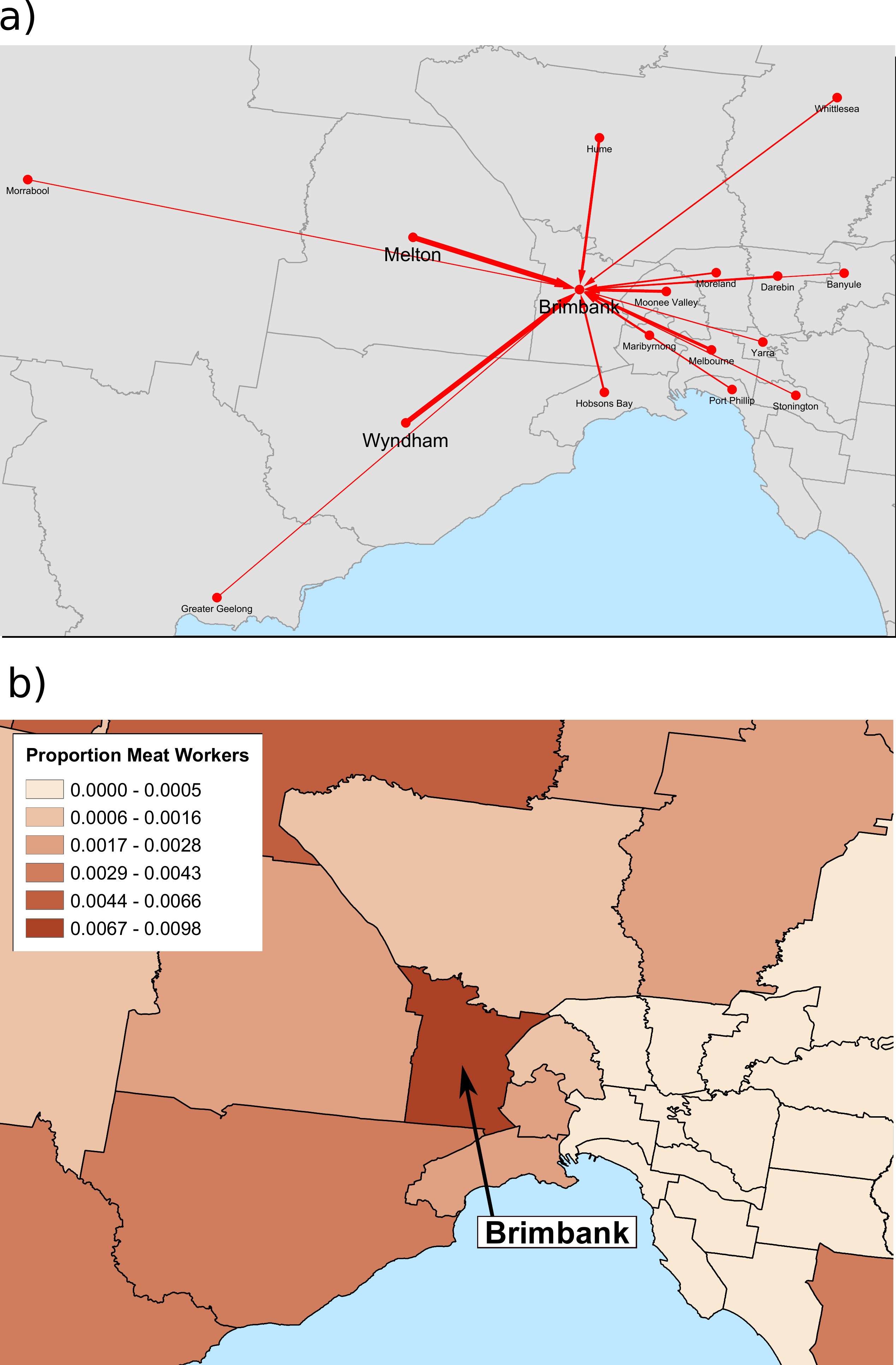}
\caption{Risk factors for spatial proliferation of SARS-CoV-2 from the Cedar Meats outbreak in Brimbank, Victoria. (a) Network visualisation of commuters into Brimbank during the period spanning April 21st to April 27th, 2020 (arrow width is proportional to commuter numbers, with the self-loop omitted). (b) The proportion of employed persons in each location working in meat processing occupations as of the 2016 Census. The colour scale in (b) was generated using the method of Jenks natural breaks.}
\label{CM_risk_factors}
\end{figure}

As our estimate of transmission risk between Brimbank and other LGAs, we compute the risk value $R_i$ as the proportion of individuals arriving in Brimbank from any other Victorian LGA $i$ during the nighttime $\rightarrow$ daytime OD matrix. These values were computed with Equation \ref{eq_point_risk} and are shown as a directed network in Figure~\ref{CM_risk_factors}a. Because the outbreak occurred in an abattoir, we also explored the effect of weighting mobility by a context-specific factor: the proportion of employed persons with occupations in meat processing (Figure \ref{CM_risk_factors}b).

\subsubsection{Risk estimates and validation with case numbers}

\begin{figure}
\centering
\includegraphics[width=0.9\columnwidth]{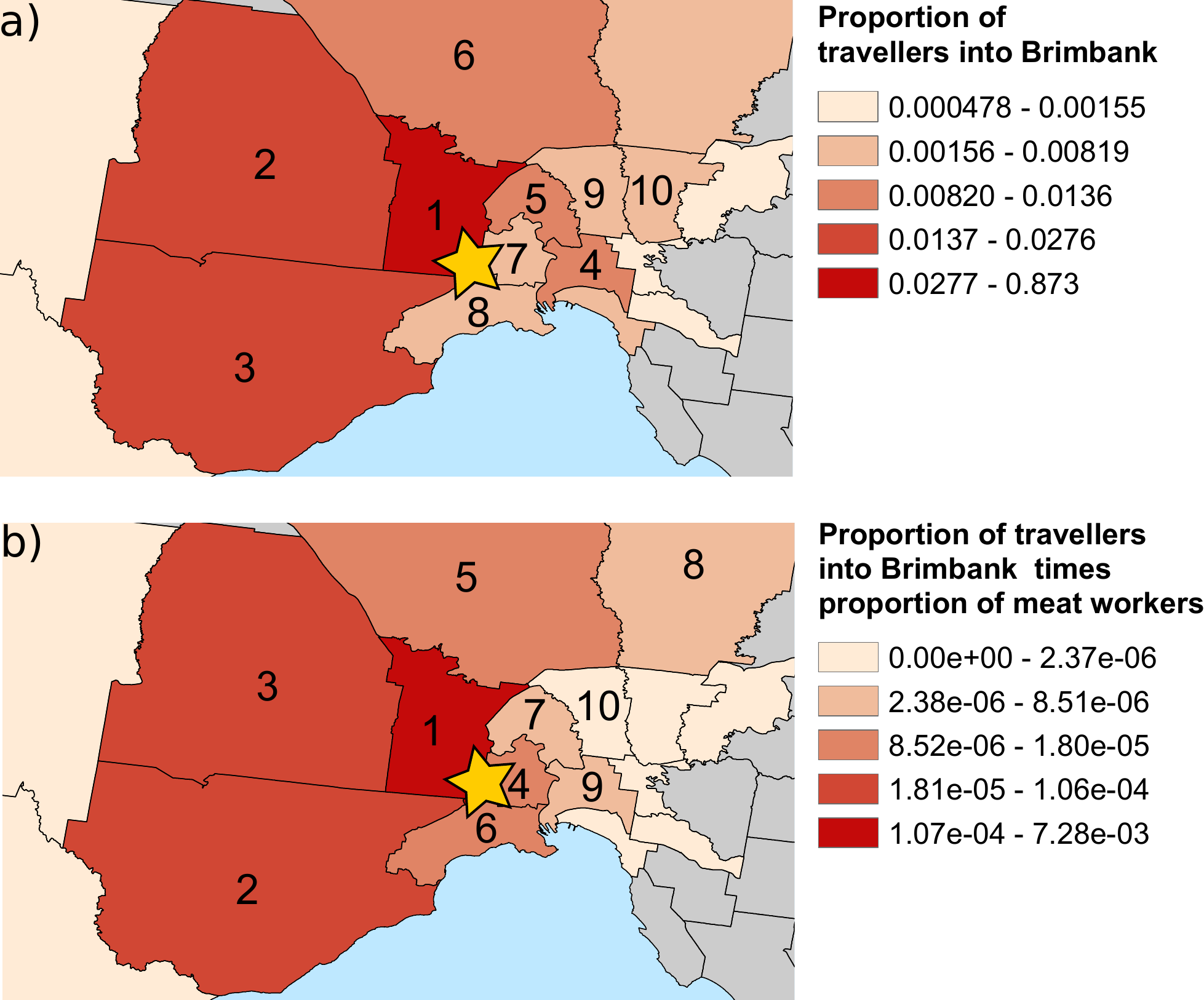}
\caption{Regional distribution of transmission risk from the Cedar Meats outbreak in Brimbank based on (a) mobility into Brimbank for daytime activities, and (b) the proportion of employed persons with occupations in meat processing multiplied by the proportion of travellers to Brimbank shown in (a). The yellow stars show the approximate location of the outbreak centre.  The numbers in (a) and (b) show the rank of each region with respect to the mapped quantity. The colour scales were generated using the method of Jenks natural breaks.}
\label{CM_risk_maps}
\end{figure}

The geographic distribution of relative transmission risk due to mobility into Brimbank during the nighttime $\rightarrow$ daytime transition is presented in Figure \ref{CM_risk_maps}(a), while the distribution generated by including both mobility and the proportion of meat workers in each LGA is shown in Figure \ref{CM_risk_maps}(b). 

\begin{figure}
\centering
\includegraphics[width=\columnwidth]{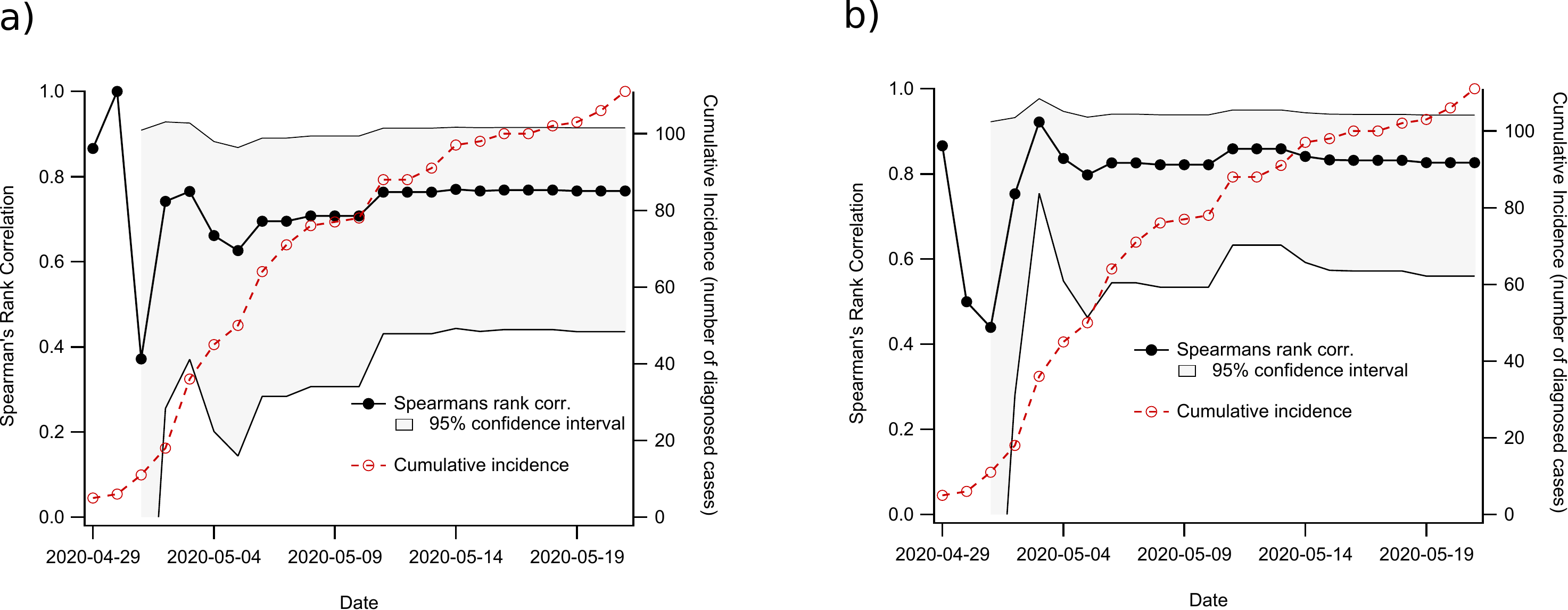}
\caption{Correlation between risk estimates and cumulative cases as a function of time. Spearman's correlation between cases by LGA and proportion of mobility into Brimbank is shown in (a), while (b) demonstrates the effect of including employment-specific contextual factors. Black dots correspond to Spearman's correlation (left y-axis), and the shaded interval is the 95\% CI. For reference, open red circles show total cumulative cases over all regions recorded for each day (right y-axis).}
\label{CM_corr}
\end{figure}

To validate our estimate, we computed Spearman's correlation between this risk estimate for each region to the time-dependent case count for each region documented over the course of the outbreak (supplied by the Victorian Department of Health and Human Services). We use Spearman's rather than Pearson's correlation because while we expect monotonic dependence between estimated relative risk and case counts, we have no reason to expect linear dependence or normally-distributed errors.
The outbreak case data was supplied as a time series of cumulative detected cases in each LGA for each day of the outbreak. Therefore, we present our correlation as a function of time from April 29th, when recorded case numbers began to increase dramatically (before May 1st, the number of affected LGAs was too small compute a confidence interval ($n\leq3$)). As case numbers increase, correlation between our risk estimates and case numbers stabilises at approximately 0.75 using mobility only (Figure \ref{CM_corr}a), and at approx. 0.81 when including both mobility and meatworker proportions in the risk computation (Figure \ref{CM_corr}b). Due to privacy limitations on release of case data, we do not present case numbers by LGA for the Cedar Meats outbreak. 

\FloatBarrier

\subsection{The Crossroads Hotel}

\subsubsection{Scenario}

The next scenario we examine began with a single spreading event that occurred during a large gathering at a social venue in Western Sydney. While workplaces have frequently been the locus of COVID-19 clusters, many outbreaks have also been sparked by social gatherings \cite{streeck2020infection,hamner2020high}. In urban environments, such outbreaks can prove more challenging to trace, as the exposed individuals may be only transiently associated with the outbreak location.

The Crossroads Hotel was the site of the first COVID-19 outbreak to occur in New South Wales after the initial wave of infections was suppressed. The cluster was identified on July 10th, 2020, during a period when new cases numbered fewer than 10 notifications per day. However, the second wave of community transmission in Victoria produced sporadic introductions in NSW, one of which led to a spreading event at the Crossroads Hotel \cite{NSW_Aug1}. Based on media reports, state contact-tracing data indicated that the cluster began on the evening of July 3rd, during a large gathering \cite{SMH_July16}.

Unlike the Cedar Meats cluster, the Crossroads Hotel scenario was not a workplace outbreak with transmission occurring in the same context for a sustained time period, but a single spreading event in a large social centre. For this reason, to estimate relevant mobility patterns we averaged trip numbers over all time-windows in our data (daytime $\rightarrow$ evening $\rightarrow$ nighttime $\rightarrow$ daytime) for the period of June 27th - July 4th. It was also necessary to perform some pre-processing of the mobility data provided by Facebook in order to correlate case data provided by New South Wales Health to our mobility-based risk estimates due to substantial differences in the geographic boundaries used in the respective data sets (see Supplementary Information and Technical Note). Aside from these minor differences, the method applied in this scenario is essentially the same as the one described above for the Cedar Meats outbreak. Risk of transmission in an area is assessed as the proportion of travellers who entered the outbreak location from that area (see Equation \ref{eq_point_risk}). 

\subsubsection{Results}

\begin{figure}
\centering
\includegraphics[width=\columnwidth]{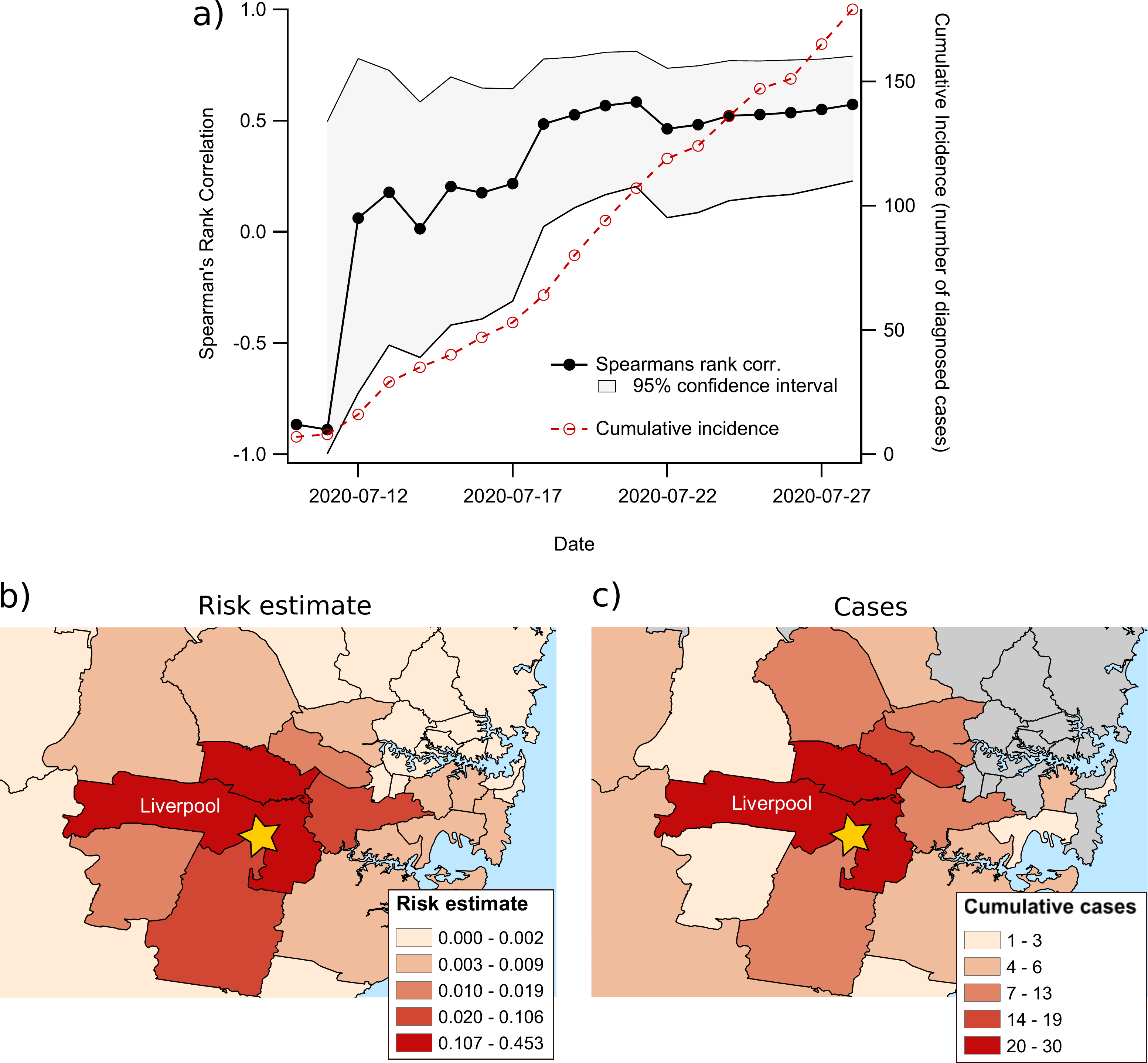}
\caption{Comparison between estimated relative risk distribution and cluster-related case numbers in New South Wales from July 12th to 28th. Spearman's rank correlation as a function of time is shown in (a). The spatial distribution of estimated relative risk computed based on mobility data recoreded for the week ending on July 4th is shown in (b), while the total number of cluster-associated cases in each LGA as of July 28th is shown in (c). The yellow star in (b) and (c) indicates the location of the outbreak centre, the Crossroads Hotel located in Liverpool, NSW. Colour scales in (b) and (c) were generated using the method of Jenks natural breaks.}
\label{CH_risk_maps}
\end{figure}

Correlation of our risk estimate to the number of cases in each LGA as a function of time is shown in Figure \ref{CH_risk_maps}(a). Heat maps of estimated risk and case numbers are shown in Figures \ref{CH_risk_maps}(b) and \ref{CH_risk_maps}(c), respectively. In this analysis, the available data did not explicitly identify the outbreak to which each case was associated, however, it did distinguish between cases associated with local transmission clusters and those associated with international importation. Because the Crossroads Hotel cluster was the only documented outbreak during this time, we attribute to it all cluster-associated cases during the period investigated. This assumption is anecdotally consistent with media reports that specify more detailed information about the residential location of individuals associated with the outbreaks. The COVID-19 case data for New South Wales is publicly available \cite{NSW_cases}. 

\subsection{Victoria: Community Transmission}

\subsubsection{Scenario}

Both of the previous case studies considered scenarios in which a localised outbreak occurred in the context of very low or undetected community transmission. For our third case study we consider a scenario in which community transmission had been detected across multiple suburbs in metropolitan Melbourne. This began in Victoria during June and July, after lifting of the social distancing restrictions that suppressed the initial outbreak in March. When community transmission was ascertained there were already active cases in many regions around metropolitan Melbourne, and no specific transmission centre was clearly identifiable (although transmission is thought to have originated in hotels used to quarantine arriving travellers).

On June 1st, 2020, Victoria DHHS reported 71 active COVID-19 cases throughout the state, with four new cases. By June 21st, this number had increased to 121 active cases (19 new) at which point the state government initiated increased testing for community transmission and re-introduction of limits on large gatherings. Subsequently, the number of active cases increased to 645 by July 6th, and localised lockdowns were implemented in a set of 12 postcodes where people were asked to stay at home unless working or attending to essential activities. These targeted lockdowns were introduced in an attempt to avoid general imposition of the measures, but they were extended to the entirety of metropolitan Melbourne on July 9th, with continuing community transmission. These events are documented in the online series of daily updates provided by Victoria DHHS \cite{DHHS_updates}. 

We examine whether the areas affected by community transmission in late June and July could have been predicted based on case numbers and mobility data that were available in early June. Our goal is to examine whether the effectiveness of mobility patterns in predicting relative transmission risk from point outbreaks can extend to community transmission scenarios in which outbreak sources are unknown. 

In the community transmission scenario, as with the Crossroads Hotel outbreak, there were no clear context-dependent factors that suggested the use of other population data. In contrast to the first two scenarios, community transmission was occurring in multiple locations at the beginning of our investigation period. For each day, the unscaled risk estimate $r_i$ is the product of the OD matrix (averaged over the preceding week) and the vector of active case numbers in each location (see Equation \ref{eq_multi_risk}). Therefore, in this case the relative risk value $R_i$ represents the proportion of travellers into all areas containing active cases, with the contribution of each infected region weighted by the number of active cases (see Equation \ref{eq_multi_risk_2}).

For this scenario, we investigate the correlation between relative risk estimates at time $t$, and incident case numbers (notifications) at time $t'$, for all dates between June 1st and July 21st. We performed this more extensive analysis because it was not clear at what point in the outbreak, if any, conditions at time $t$ would provide insight at a future time $t'$. In particular, we investigate if and when the incorporation of mobility data gives insight not provided by active case numbers alone.

\subsubsection{Results}
\begin{figure}
\centering
\includegraphics[width=\columnwidth]{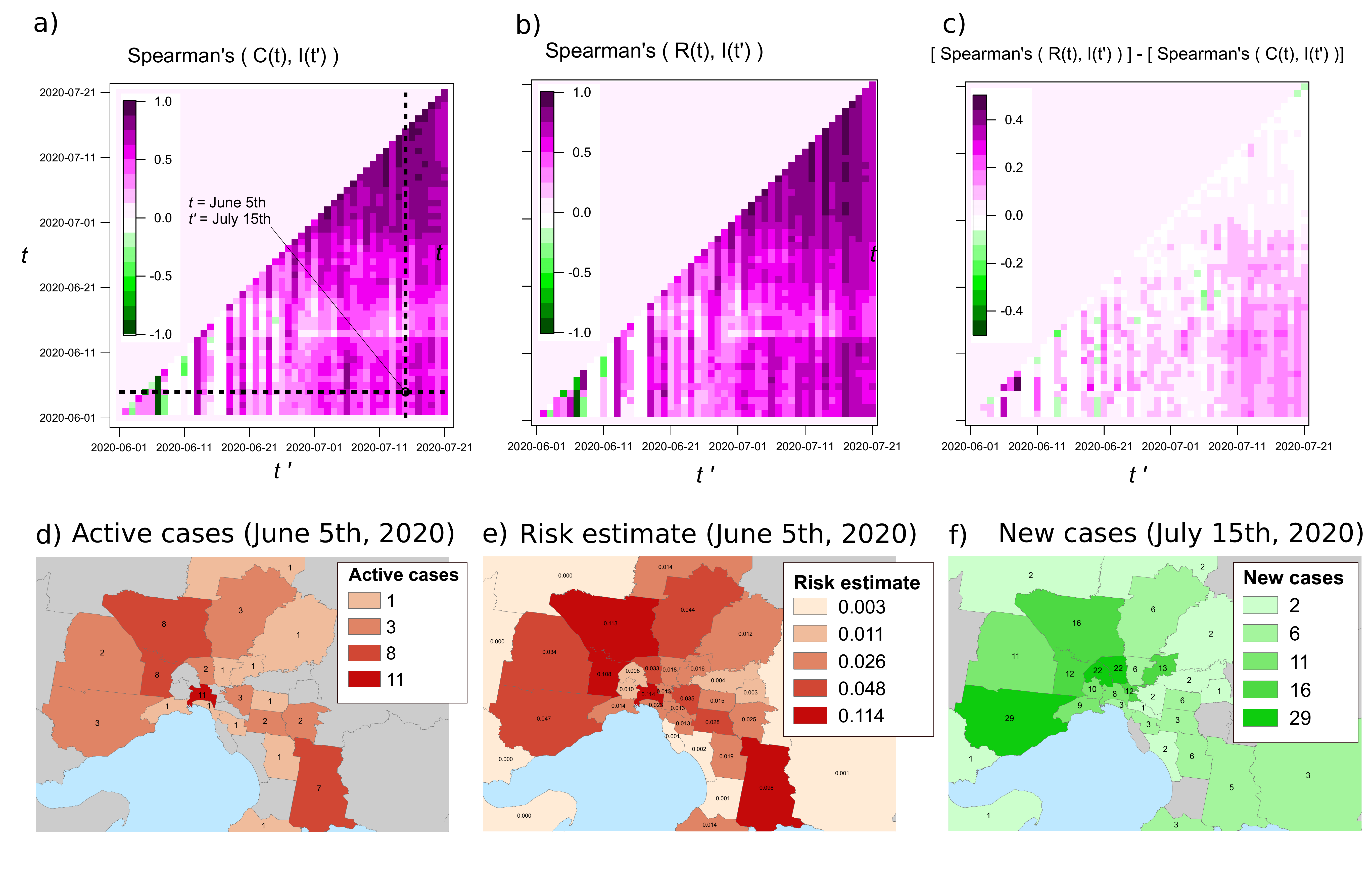}
\caption{The contribution of mobility information to relative risk estimates in a community transmission scenario. The correlation between active cases at time $t$ and incident cases at time $t'$ is shown in (a) while the correlation of the mobility-based relative risk estimation at time $t$ with incident cases at time $t'$ is shown in (b). The benefit of including mobility information is indicated in (c), which shows the values plotted in (b) minus those plotted in (a). The maps in (d), (e), and (f) correspond to the $(t, t')$ point indicated in (a), and show the number of active cases on June 6th (d), the distribution of mobility-related relative risk on that day (e), and the number of incident cases on July 15th (f). The colour scales in (d), (e), and (f) were generated using the method of Jenks natural breaks.}
\label{CommTrans}
\end{figure}

The results of our correlation analysis for the Victoria community transmission scenario are shown in Figure \ref{CommTrans}. The correlations of incident cases ($I$) at time $t'$ with active case numbers ($C$) and active cases combined with mobility ($R$) at time $t$ are shown in Figures \ref{CommTrans}(a) and \ref{CommTrans}(b), respectively.

The added contribution of the mobility data as a function of $t$ and $t'$ is shown in Figure \ref{CommTrans}(c), which shows the difference between the mobility-based correlation value and the correlation based on active case numbers alone. 

The values in Figure \ref{CommTrans}(c) test the hypothesis that, after some delay, risk estimates incorporating mobility will correspond more closely to the future distribution of infection incidence than estimates made based only on active case numbers. Positive values support this hypothesis while values near or below zero correspond to the null hypothesis that mobility information does not improve risk estimates based on the distribution of active cases. Furthermore, by examining these values as functions of the time of risk assessment ($t$) and the time of case reports ($t'$), we gain some insight into the temporal lag between the mobility-driven diffusion of disease and the official reporting of cases. Here, the lag $(t' - t)$ between observation of risk at time $t$ and observation of case incidence at time $t'$ integrates all sources of delay. These include both the natural period between transmission and symptom onset, as well as the logistical delays associated with clinical presentation, testing, ascertainment, and notification. Our simple analysis does not allow us to decompose the dynamics to assess different lag components separately.

To demonstrate the geographic distribution of cases and the diffusion of risk based on mobility, Figure \ref{CommTrans}(d) shows the active case counts documented for June 5th, Figure \ref{CommTrans}(e) shows the corresponding distribution of transmission risk based on mobility patterns from the preceding week, and Figure \ref{CommTrans}(f) shows the distribution of incident cases on July 15th. For reference, the maps in Figures \ref{CommTrans}(d), \ref{CommTrans}(e), and \ref{CommTrans}(f) correspond to the point indicated by the intersection of dashed lines in Figure \ref{CommTrans}(a).

\FloatBarrier

\section{Discussion}

The goal of this study was to develop and critically analyse a simple procedure for translating aggregate mobility data into estimates of the spatial distribution of relative transmission risk from COVID-19 outbreaks. Our results indicate that aggregate mobility data can be a useful tool in estimation of COVID-19 transmission risk diffusion from locations where active cases have been identified. The utility of mobility data depends on the context of the outbreak and appears to be more helpful in scenarios involving environments where context indicates specific risk factors. The procedure we presented may also be useful during the early stages of community transmission and could help determine the extent of selective intervention measures.

In community transmission scenarios, mobility will already have played a role in determining the distribution of case counts when community transmission is detected. Our results indicate that the insight added by the incorporation of mobility data diminishes as case counts grow. However, we also observed low correlations due to stochastic effects in the Crossroads Hotel scenario. Taken together, these results indicate that there is an optimal usage window that opens when case counts are high enough for aggregate mobility patterns to shed light on  transmission patterns, and closes when these transmission patterns begin to determine the distribution of active cases which then predict their own future distribution with only limited information added by considering mobility. In addition, once case counts rise sufficiently to trigger intervention policies, the local stringency of these measures and adherence to them are likely to become primary factors influencing the incidence of new cases \cite{sen2020association,kraemer2020effect}.

Our examination of the second wave of community transmission in Victoria showed that several weeks before it was recognised, the spatial distribution of a small number of active cases was indicative of the outbreak distribution more than 30 days later when interventions were introduced. This indication improved slightly by including the diffusion of risk computed from available mobility data. Qualitatively, this observation indicates that even when case numbers were small, low-level community transmission may have already been taking place throughout the region of metropolitan Melbourne. This suggests that earlier selective lockdown measures, extending beyond the borders of regions in which cases had been identified, may have been more effective at containing transmission.

This type of relative risk estimation procedure is relevant to public health decisions relating to selective lockdown measures or the imposition of mandated infection control policies upon either the initial introduction of an infectious disease into a susceptible population or the resurgence of a previously suppressed epidemic. Australia is currently (as of August, 2020) in the early phases of the latter scenario and there is a need for policy decision frameworks aimed at preventing resurgence of the epidemic while minimising economic consequences of further intervention. Importantly, this study focused on relatively small-scale outbreaks that all occurred within single administrative jurisdictions. For scenarios involving case importation between different administrative jurisdictions (i.e., international or interstate travel), calculation of transmission risk must take into account heterogeneity in the rates of case ascertainment within each jurisdiction, as these can vary substantially \cite{niehus2020using}.

\subsection{Limitations}

\subsubsection{Privacy, anonymity, and aggregation}

It is essential that the use of mobility data for disease surveillance comply with privacy and ethical considerations \cite{buckee2020aggregated}. Due to this requirement, there will always be trade-offs between the spatiotemporal resolution of aggregated mobility data and the completeness of the data set after curation, which typically involves the addition of noise and the removal of small numbers based on a specified threshold. To help ensure users cannot be identified, Facebook removes OD pairs with fewer than 10 unique users over the 8-hr aggregation period. The combination of this aggregation period with the 10-user threshold affects regional representation in the data set, particularly in more sparsely populated areas. The final product resulting from these choices contains frequently-updated and temporally-specific mobility patterns for densely populated urban areas, at the cost of incomplete data in sparsely populated regions. In general, increased temporal or spatial resolution will reduce trip numbers in any given set of raw data, which can have a dramatic impact on the amount of information missing from the curated numbers \cite{fair2019creating}.

\subsubsection{Stochastic effects}
The comparison of our results from the Cedar Meats outbreak and those from the Crossroads Hotel cluster demonstrate that the utility of aggregated mobility patterns in estimation of the spatial distribution of relative risk depends on the context of the outbreak, with more value in situations involving habitual mobility such as commuting to and from work. Detailed examination of the inconsistencies between risk estimates and case data from the Crossroads Hotel outbreak indicate that small numbers of people travelling longer distances were responsible for the relative lack of correspondence in that scenario. In particular, news reports discussed instances of single individuals who had travelled from the rural suburbs to visit the Crossroads Hotel for the July 3rd gathering who then infected their family members. These scenarios were not consistent with the risk predictions produced by the mobility patterns into and out of the region and exemplify the limitations of risk assessment based on aggregate behavioural data. 

\subsubsection{Sample bias}
The mobility data provided by the Facebook Data For Good program represents a non-uniform and essentially uncharacterised sample of the population. While it is a large sample, with aggregate counts on the order of $10\%$ of ABS population figures, the spatial bias introduced by the condition of mobile app usage cannot be determined due to data aggregation and anonymisation. While it is possible to count the number of Facebook users present in any location during the specified time-intervals, it is not possible to distinguish which of those are located in their places of residence. In order to account for the (possibly many) biases affecting the sample, a detailed demographic study would be necessary that is beyond the scope of the present work. A heat map (Supplementary Figure \ref{FB_pop}) of the average number of Facebook users present during the nighttime period (2am to 10am) as a proportion of the estimated resident population reported by the ABS (2018 \cite{ABS_pop18}) shows qualitative similarity to the spatial distributions of active cases and relative risk shown in Figure \ref{CommTrans}(d) and (e). This dual-correspondence suggests the presence of common factors affecting both representation in the Facebook dataset and the risk of transmission. To investigate the potential influence of spatial sampling bias on our correlations, we performed a simple bias correction the results of which are shown in Supplementary Figure \ref{corrected}. We did not include this bias correction as a component of our general analysis because it is unclear to what degree the correction is accurate, given a lack of detailed information on the individuals represented in Facebook user population data. That is, the bias correction we tested may have introduced different, uncharacterised biases.

\subsection{Future Work}
On a fundamental level, mobility patterns are responsible for observed departures from continuum mechanics observed in real epidemics \cite{viboud2006synchrony}.  Over the past two decades, due to public health concern over the pandemic potential of SARS, MERS, and novel influenza, spatially explicit models of disease transmission have become commonplace in simulations of realistic pandemic intervention policies \cite{Germann5935,zachreson2020interfering}. Such models rely on descriptions of mobility patterns which are usually derived from static snapshots of mobility obtained from census data \cite{moss2019can,cliff2018investigating,fair2019creating}. While this approach is justifiable given the known importance of mobility in disease transmission, it is also clear that the shocks to normal mobility behaviour induced by the intervention policies of the COVID-19 pandemic will not be captured by static treatments of mobility patterns. To account for the dynamic effects of intervention, several models have been developed to simulate the imposition of social distancing measures through adjustments to the strength of context-specific transmission factors \cite{ferguson2020report,chang2020modelling}. This type of treatment implicitly affects the degree of mixing between regions without explicitly altering the topology of the mobility network on which the model is based and it is unclear whether such a treatment is adequate to capture the complex response of human population behaviour. Given the results of our analysis, the incorporation of real-time changes in mobility patterns could add policy-relevant layers of realism to such models that currently rely on static, sometimes dated, depictions of human movement. 

\section{Code and Data Availability}

Example scripts and data used for computing risk estimates and correlations can be found in the associated GitHub repository:

\url{https://github.com/cjzachreson/COVID-19-Mobility-Risk-Mapping}

However, due to release restrictions on the mobility data provided by Facebook, the OD matrices are not included as these were derived from the data provided by the Facebook Data For Good program (random matrices are included as placeholders). The processed mobility data used in this work may be made available upon request to the authors, subject to conditions of release consistent with the Facebook Data For Good Program access agreement.  

A generic implementation of the code used to re-partition OD matrices between different geospatial boundary definitions is enclosed in the supplementary Technical Note. 

\section{Acknowledgements}

The authors would like to acknowledge useful discussions with Nick Golding, Freya Shearer, Jodie McVernon, James McCaw, and James Wood. We thank the Facebook Data For Good Program and in particular would like to acknowledge the continuous efforts of Alex Pompe in data provision and technical discussion.  

\section{Author Contributions}

All authors contributed to manuscript composition, research design, and discussion. CZ, NG, NR, and LM designed the risk estimation procedure. CZ and MJL performed validation of risk estimates against case data. CZ, NR, and MT performed spatial data processing. CZ implemented data analysis scripts, and composed figures.

\section{Funding Statement}
 This work was supported in part by and NHMRC project grant (APP1165876), an NHMRC Centre of Research Excellence (APP1170960), and the Victoria State Government Department of Health and Human Services.

\FloatBarrier
\newpage

\newcommand{\beginsupplement}{%

 \setcounter{table}{0}
   \renewcommand{\thetable}{S\arabic{table}}%
   
     \setcounter{figure}{0}
      \renewcommand{\thefigure}{S\arabic{figure}}%
      
      \setcounter{page}{1}
      \renewcommand{\thepage}{S\arabic{page}} 
      
      \setcounter{section}{0}
      \renewcommand{\thesection}{S\arabic{section}}
      
      \setcounter{equation}{0}
      \renewcommand{\theequation}{S\arabic{equation}}
     }


\beginsupplement
\begin{center}
\Large{\bf{Supplementary Information for: \\ Risk mapping for COVID-19 outbreaks in Australia using mobility data}}\\
\vspace{0.5cm}
\large{Cameron Zachreson, Lewis Mitchell, Michael J. Lydeamore, Nicolas Rebuli, Martin Tomko, Nicholas Geard}
\end{center}
\section{Description of Mobility Data}

The data used in our study was provided by the Facebook Data for Good program. The data set (in the Disease Prevention Maps subset) is aggregated from individual-level GPS coordinates collected from the use of Facebook's mobile app. Therefore, the raw data is biased to over-represent the movements of any subpopulations more likely to utilise social media applications on mobile devices. After collection, the data is spatially and temporally aggregated as a list of trip numbers between Bing Tiles \cite{BingTiles} within a rectangular raster pattern (i.e., centered on a country, state, or city). The sizes and boundaries of these discrete locations are determined by an optimisation procedure that produces the smallest subregion size possible (down to a minimum size of 600m $\times$ 600m), given the extent of the region of interest and the requirement for near-real time release of new data. A trip between locations is defined based on the most frequently visited tile in the first 8-hour period and the most frequently-visited tile in the subsequent 8-hour period. Finally, before the data is released, any entries showing fewer than 10 trips between a pair of locations are removed to protect the privacy of individual users. For Australia, the state-level data consists of trip numbers between 2km $\times$ 2km tiles. By comparing this scale to larger (national-scale) and smaller (city-scale) regions of interest, we determined that the state-level data provided the best balance, with trip numbers large enough to produce a sufficiently dense network of connections while still providing a subregion size that is usually smaller than the Local Government Areas for which case data is reported.

\subsection{Generating correspondences}
Because the raw mobility data is provided as movements between tiles, while case data is provided based on the boundaries of Local Government Areas. We note that while Facebook releases data aggregated to administrative regions, these regions were not geographically consistent with the current LGA boundaries for Australia. In order to ensure consistency of our method across datasets and jurisdictions, we produced our own correspondence system. We did this by performing two spatial join operations. These associate either tiles or LGAs with Meshblocks (the smallest geographic partition on which the Australian Bureau of Statistics releases population data). Meshblocks were associated based on their centroid locations. Each meshblock centroid was associated to the tile with the nearest centroid and to the LGA containing it. We did not split meshblocks whose boundaries lay on either side of an LGA or tile boundary, as their sizes are sufficiently small that edge effects are negligible (in addition, the set of LGAs forms a complete partition of meshblocks, so edge effects were only observed for tile associations). We then associated tiles to LGAs proportionately based on the fraction of the total meshblock population within that tile that was associated with each overlapping LGA. 

\subsection{Re-partitioning mobility data}
Once a correspondence is established between the tile partitions on which mobility data is released and the LGA partitions on which case data is released, the matrix of connections between tiles must be converted into a matrix of connections between LGAs. The Supplementary Technical Note explains how we performed this step, and gives a general method for converting matrices between partition schemes. Briefly, the number of trips between two locations in the initial data is split between the overlapping set of partitions in the new set of boundaries (in this case, local government areas), based on the correspondence between partition schemes determined as explained in the previous subsection. 

\subsection{Spatial biases in Facebook mobility data}

To investigate the spatial sample biases present in the mobility data provided by Facebook, we examined the ratio of Facebook users to ABS 2018 population for each suburb in Victoria. While the true number varies from day to day, an example of this distribution is shown as a heat map in Supplemental Figure \ref{FB_pop}, which displays the average number of Facebook mobile app users indexed to each LGA between the hours of 2am and 10am from May 15th to June 25th, divided by the estimated resident population reported by the ABS in 2018. The distribution is narrow, with most urban areas falling in the range of 5 \% to 10 \% Facebook users. However, this is not an exact representation of residential population proportions, as many mobile users work during the nighttime and will not be located at their residence during the selected period. Unfortunately, it is not possible to precisely quantify the bias introduced by Facebook's sampling scheme. 

Despite these limitations, it may still be informative to examine whether accounting for the bias pictured in Figure \ref{FB_pop} affects our validation. To determine this, we re-computed the correlations pictured in Figures \ref{CM_corr}(a) and \ref{CommTrans}(b) (corresponding to the Cedar Meats and Victoria community transmission scenarios). To do so, we multiplied all mobility flows out of each region by the inverse proportion of Facebook users to the total number of residents in the origin location. For the reasons discussed above, this is not an exact accounting for sample bias, but may partially correct for heterogeneity in the proportion of travellers counted in mobility data released by Facebook.

\begin{figure}
\centering
\includegraphics[width=\columnwidth]{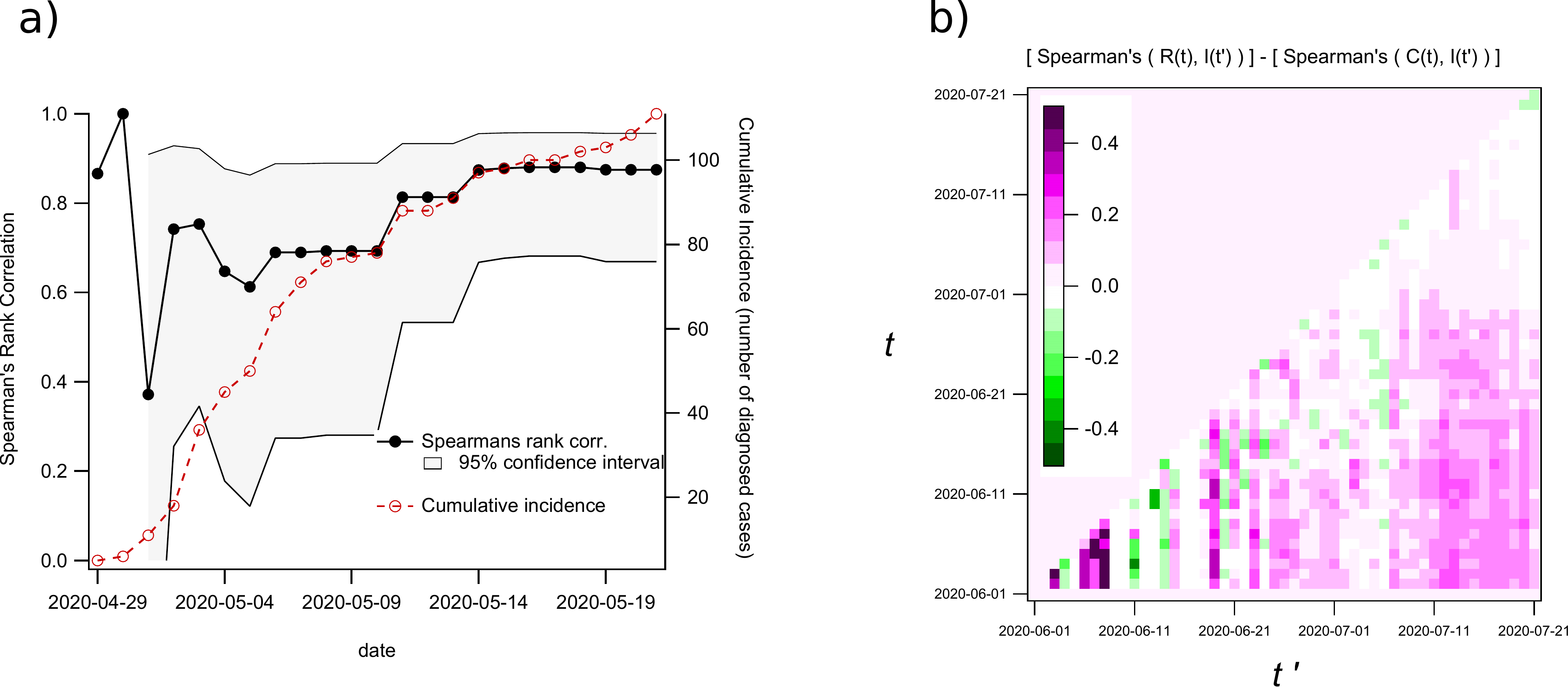}
\caption{Correlations between partially bias-corrected risk estimates and documented case numbers. Spearman's rank correlation between cumulative case numbers by LGA and the modified relative risk estimate for the Cedar Meats outbreak is shown in (a). Subfigure (b) shows the partially bias-corrected version of Figure \ref{CommTrans}(c), demonstrating the effect of including bias-corrected mobility with active case numbers at time $t$ in estimation of relative incident case risk at time $t'$.}
\label{corrected}
\end{figure}

For the Cedar Meats outbreak scenario, accounting for the Facebook sample bias in this way improves the correlation between our mobility-based relative risk estimate and the recorded case counts (Figure \ref{corrected}a). For the community transmission scenario, performing this extra step does not appear to substantially change the result shown in Figure \ref{CommTrans} (compare Figure \ref{corrected}b and Figure \ref{CommTrans}c). 

\begin{figure}
\centering
\includegraphics[width=\columnwidth]{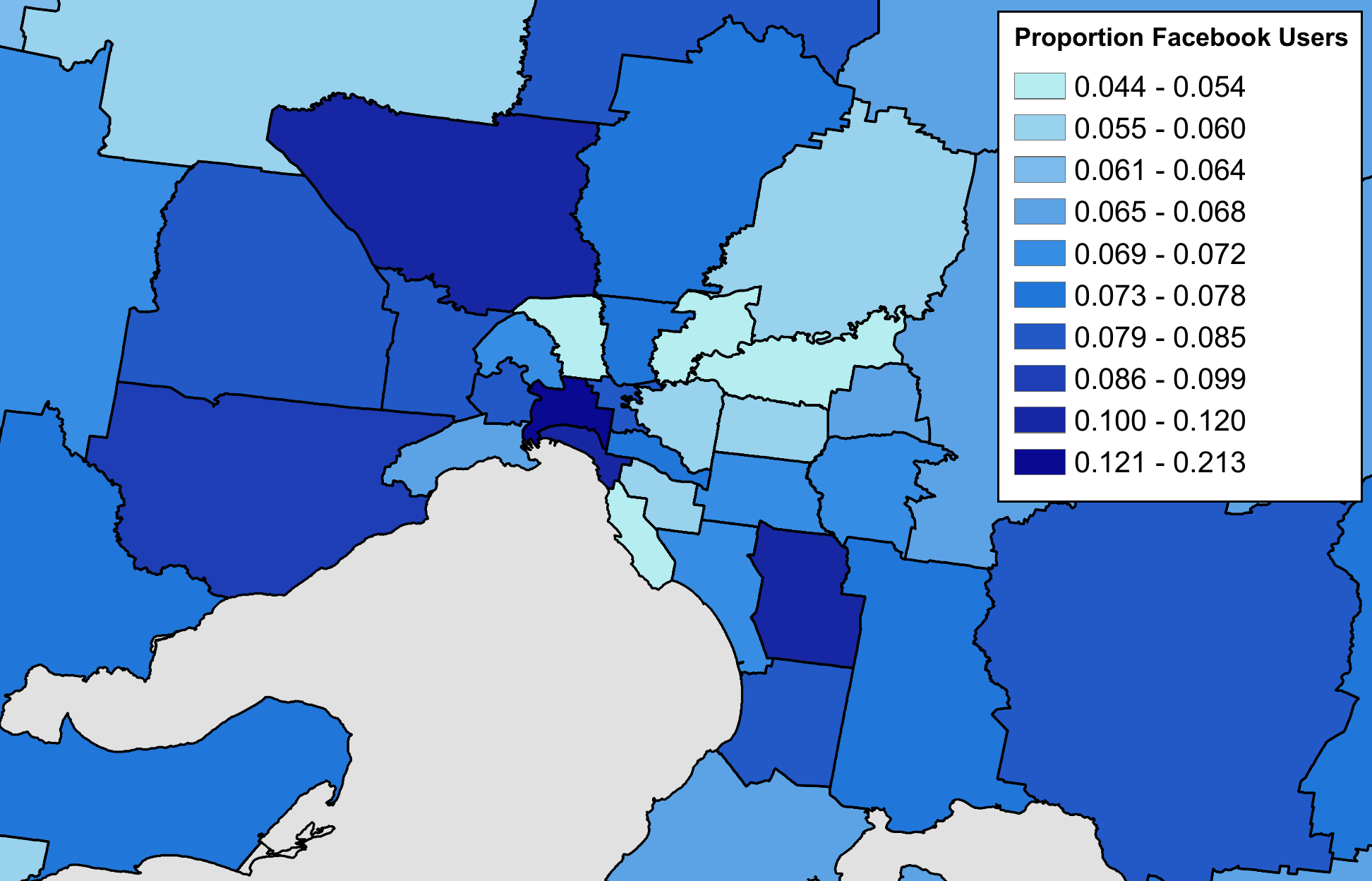}
\caption{Heat map of the proportion of Facebook users estimated for each LGA. The values are computed as the number of Facebook users who were found in each location during the hours of 2am until 10am averaged from May 15th to June 25th, divided by the residential population recorded by the ABS in 2018. The colour scale was generated using the method of Jenks natural breaks. }
\label{FB_pop}
\end{figure} 

\subsection{Temporal autocorrelation of mobility matrices}

\begin{figure}
\centering
\includegraphics[width=\columnwidth]{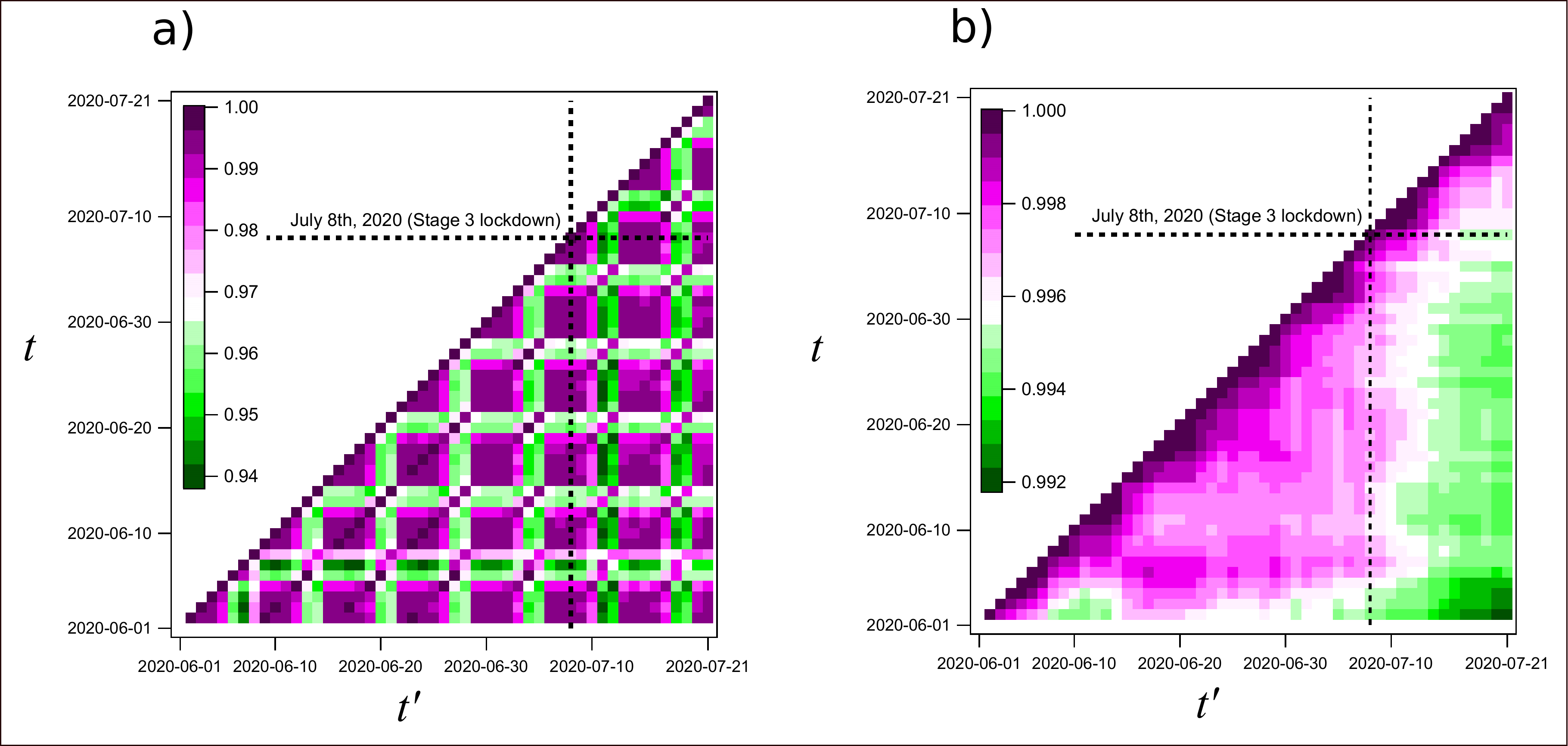}
\caption{Autocorrelation (Spearman's rho) of mobility flows at time $t$ (y-axis) and $t'$ (x-axis), for all LGA origin-destination pairs in Victoria, between June 1st and July 21st, 2020. Subfigure (a) shows the raw autocorrelation values, and (b) shows the autocorrelation of the 7-day average. The black dashed lines indicate the date on which Stage 3 distancing restrictions were implemented in the regions of Greater Melbourne and Mitchell Shire. }
\label{OD_autocorr}
\end{figure}

In order to investigate the degree to which mobility changed during the study period, we computed the autocorrelation of mobility flows between origin-destination pairs at time $t$ to those at future times $t'$. The results, shown in Figure \ref{OD_autocorr}, demonstrate that while weekend and weekday mobility differ markedly, and the implementation of stage 3 restrictions in Greater Melbourne altered mobility patterns, there is a very high level of temporal consistency in relative mobility volumes throughout the studied period. For this reason, our results for the community transmission scenario shown in Figure \ref{CommTrans} are robust to the precise selection of time periods used to generate the mobility matrices for our risk estimates. For example, if we integrate mobility flows over a period longer than one week, or consider a nonzero delay between the period over which mobility is averaged and the time $t$ for which active case data is tabulated, it has no effect on the resulting risk rankings and gives the same pattern of temporal correlations (though the risk values themselves are affected slightly). 

\section{Correlating Risk Estimates to Case Data} 

We used Spearman's rank correlation to investigate the correspondence between our relative risk estimates and documented case data. This measure of correlation is typically used when comparing ordinal data, or, more generally, when monotonic relationships are expected, but errors are not normally-distributed. In order to investigate the monotonicity between relative risk estimates and reported case numbers, we aligned the documented case data for all regions in which infections had been tabulated against the corresponding relative risk estimates for those regions. Note that our correlations did not include regions for which no case data was available. Therefore, our correlation results illustrate the degree to which risk estimates are monotonic with case numbers, but do not account for any risk estimates made in areas with no cases to compare to. This results in a high degree of uncertainty when the number of affected areas is small, reflected by the wide confidence intervals observed in the early stages of the Cedar Meats and Crossroads Hotel outbreaks (Figures \ref{CM_corr}, and \ref{CH_risk_maps}a, respectively). 

The 95\% confidence intervals were computed using Fisher's Z transformation with quantile parameter $\alpha = 1.96$.

\section{ABS Data Sources}

Two data sets from the Australian Bureau of Statistics were used in this study: 1) number of residents by industry of occupation (2016), and 2) resident population (2018).

\subsection{Population by LGA}

The distributions shown in Figure \ref{FB_pop} were computed by dividing the number of Facebook users indexed to each LGA during the nighttime period by the resident population in each LGA. We obtained the population data from the ABS 2018 population dataset which is publicly available \cite{ABS_pop18}. The Facebook user populations are provided by the Data For Good program in addition to the mobility data discussed above.  

\subsection{Employed persons by industry of occupation}

As a context-specific risk factor for the Cedar Meats outbreak we obtained the number of individuals by place of usual residence and industry of occupation. Specifically, we obtained the number of residents in each Local Government Area (2016 boundaries), employed in the occupation categories ``Meat Boners and Slicers and Slaughterers" and ``Meat Poultry and Seafood Process Workers". This data from the 2016 Australian Census of Population and Housing is available from the ABS TableBuilder web application \cite{ABS_TB}. We used a population-weighted correspondence list to convert the data provided on geospatial boundaries of 2016 Local Government Areas into 2018 Local Government Area boundaries. For the Melbourne region in which this data was applied, these boundaries have not changed substantially between 2016 and 2018.

To compute the factors used to weight the mobility-based relative risk predictions, we divided the total number of workers in both of the above categories by the number of employed persons (those employed full time or part-time) in each LGA, which we also drew from the 2016 Australian Census via Census TableBuilder.

\section{Case Data} 

COVID-19 case data by local government area is available from Australian jurisdictional health authorities. For this work, we used data provided by NSW Health \cite{NSW_cases} (all data is publicly available) and from Victoria DHHS. The data used for the Cedar Meats outbreak scenario was obtained from DHHS through a formal request to the Victorian Agency for Health Information (VAHI) and cannot be made public in this work. The case data by LGA used to evaluate the Victoria community transmission scenario was taken directly from the COVID-19 daily update archives available on the DHHS public website \cite{DHHS_updates}.

\section{Description of Supplemental Data}

\begin{itemize}
\item{Timeseries of total case incidence for the Crossroads Hotel and Cedar Meatworks studies}
\item{Correlation values used in Figure \ref{CommTrans}(a), (b), and (c)}
\item{95\% confidence intervals for Figures \ref{CommTrans}(a) and \ref{CommTrans}(b)}
\end{itemize}

\section{References}

\begin{thebibliography}{10}

\bibitem{kutter2018transmission}
Kutter JS, Spronken MI, Fraaij PL, Fouchier RA, Herfst S.
\newblock Transmission routes of respiratory viruses among humans.
\newblock Curr Opin Virol. 2018;28:142--151.

\bibitem{siegel20072007}
Siegel JD, Rhinehart E, Jackson M, Chiarello L, Committee HCICPA, et~al.
\newblock 2007 guideline for isolation precautions: preventing transmission of
  infectious agents in health care settings.
\newblock Am J Infect Control. 2007;35(10):S65.

\bibitem{lauer2020incubation}
Lauer SA, Grantz KH, Bi Q, Jones FK, Zheng Q, Meredith HR, et~al.
\newblock The incubation period of coronavirus disease 2019 (COVID-19) from
  publicly reported confirmed cases: estimation and application.
\newblock Ann Intern Med. 2020;172(9):577--582.

\bibitem{he2020temporal}
He X, Lau EH, Wu P, Deng X, Wang J, Hao X, et~al.
\newblock Temporal dynamics in viral shedding and transmissibility of COVID-19.
\newblock Nat Med. 2020;26(5):672--675.

\bibitem{young2020epidemiologic}
Young BE, Ong SWX, Kalimuddin S, Low JG, Tan SY, Loh J, et~al.
\newblock Epidemiologic features and clinical course of patients infected with
  SARS-CoV-2 in Singapore.
\newblock JAMA. 2020;323(15):1488--1494.

\bibitem{ferretti2020quantifying}
Ferretti L, Wymant C, Kendall M, Zhao L, Nurtay A, Abeler-D{\"o}rner L, et~al.
\newblock Quantifying SARS-CoV-2 transmission suggests epidemic control with
  digital contact tracing.
\newblock Science. 2020;368(6491).

\bibitem{li2020substantial}
Li R, Pei S, Chen B, Song Y, Zhang T, Yang W, et~al.
\newblock Substantial undocumented infection facilitates the rapid
  dissemination of novel coronavirus (SARS-CoV-2).
\newblock Science. 2020;368(6490):489--493.

\bibitem{wei2020presymptomatic}
Wei WE, Li Z, Chiew CJ, Yong SE, Toh MP, Lee VJ.
\newblock Presymptomatic Transmission of SARS-CoV-2—Singapore, January
  23--March 16, 2020.
\newblock Morbidity and Mortality Weekly Report. 2020;69(14):411.

\bibitem{arons2020presymptomatic}
Arons MM, Hatfield KM, Reddy SC, Kimball A, James A, Jacobs JR, et~al.
\newblock Presymptomatic SARS-CoV-2 infections and transmission in a skilled
  nursing facility.
\newblock N Engl J Med. 2020;.

\bibitem{kimball2020asymptomatic}
Kimball A, Hatfield KM, Arons M, James A, Taylor J, Spicer K, et~al.
\newblock Asymptomatic and presymptomatic SARS-CoV-2 infections in residents of
  a long-term care skilled nursing facility—King County, Washington, March
  2020.
\newblock Morbidity and Mortality Weekly Report. 2020;69(13):377.

\bibitem{buckee2020aggregated}
Buckee CO, Balsari S, Chan J, Crosas M, Dominici F, Gasser U, et~al.
\newblock Aggregated mobility data could help fight COVID-19.
\newblock Science. 2020;368(6487):145.

\bibitem{pepe2020covid}
Pepe E, Bajardi P, Gauvin L, Privitera F, Lake B, Cattuto C, et~al.
\newblock COVID-19 outbreak response, a dataset to assess mobility changes in
  Italy following national lockdown.
\newblock Sci Data. 2020;7(1):1--7.

\bibitem{martin2020effectiveness}
Mart{\'\i}n-Calvo D, Aleta A, Pentland A, Moreno Y, Moro E.
\newblock Effectiveness of social distancing strategies for protecting a
  community from a pandemic with a data driven contact network based on census
  and real-world mobility data.
\newblock In: Technical Report; 2020. .

\bibitem{bourassa2020social}
Bourassa K, Sbarra D, Caspi A, Moffitt T.
\newblock Social Distancing as a Health Behavior: County-level Movement in the
  United States During the COVID-19 Pandemic is Associated with Conventional
  Health Behaviors. 2020;.

\bibitem{maas2019facebook}
Maas P, Iyer S, Gros A, Park W, McGorman L, Nayak C, et~al.
\newblock Facebook Disaster Maps: Aggregate Insights for Crisis Response \&
  Recovery.
\newblock In: ISCRAM; 2019. .

\bibitem{bonaccorsi2020economic}
Bonaccorsi G, Pierri F, Cinelli M, Flori A, Galeazzi A, Porcelli F, et~al.
\newblock Economic and social consequences of human mobility restrictions under
  COVID-19.
\newblock Proc Natl Acad Sci USA. 2020;117(27):15530--15535.

\bibitem{lee2020job}
Lee K, Sahai H, Baylis P, Greenstone M.
\newblock Job Loss and Behavioral Change: The Unprecedented Effects of the
  India Lockdown in Delhi.
\newblock University of Chicago, Becker Friedman Institute for Economics
  Working Paper. 2020;(2020-65).

\bibitem{holtz2020interdependence}
Holtz D, Zhao M, Benzell SG, Cao CY, Rahimian MA, Yang J, et~al.
\newblock Interdependence and the Cost of Uncoordinated Responses to COVID-19.
  2020;.

\bibitem{galeazzi2020human}
Galeazzi A, Cinelli M, Bonaccorsi G, Pierri F, Schmidt AL, Scala A, et~al.
\newblock Human Mobility in Response to COVID-19 in France, Italy and UK.
\newblock arXiv preprint arXiv:200506341. 2020;.

\bibitem{kraemer2020effect}
Kraemer MU, Yang CH, Gutierrez B, Wu CH, Klein B, Pigott DM, et~al.
\newblock The effect of human mobility and control measures on the COVID-19
  epidemic in China.
\newblock Science. 2020;368(6490):493--497.

\bibitem{jia2020population}
Jia JS, Lu X, Yuan Y, Xu G, Jia J, Christakis NA.
\newblock Population flow drives spatio-temporal distribution of COVID-19 in
  China.
\newblock Nature. 2020;p. 1--5.

\bibitem{niehus2020using}
Niehus R, De~Salazar PM, Taylor AR, Lipsitch M.
\newblock Using observational data to quantify bias of traveller-derived
  COVID-19 prevalence estimates in Wuhan, China.
\newblock The Lancet Infectious Diseases. 2020;.

\bibitem{Aus_COVID}
Coronavirus (COVID-19) at a glance – 10 August 2020; 2020.
\newblock [Online; accessed 13-Aug-2020].
\newblock
  \url{https://www.health.gov.au/resources/publications/coronavirus-covid-19-at-a-glance-10-august-2020}.

\bibitem{dyal2020covid}
Dyal JW.
\newblock COVID-19 Among Workers in Meat and Poultry Processing Facilities―19
  States, April 2020.
\newblock MMWR Morbidity and mortality weekly report. 2020;69.

\bibitem{richmond2020interregional}
Richmond CS, Sabin AP, Jobe DA, Lovrich SD, Kenny PA.
\newblock Interregional SARS-CoV-2 spread from a single introduction outbreak
  in a meat-packing plant in northeast Iowa.
\newblock medRxiv. 2020;.

\bibitem{BeefCentral}
Sim T. First Cedar Meats COVID-19 case confirmed on 2 April; 2020.
\newblock [Online; accessed 11-Aug-2020].
\newblock
  \url{https://www.beefcentral.com/processing/first-cedar-meats-covid-19-case-confirmed-on-2-april/}.

\bibitem{DHHS_May2}
Coronavirus update for Victoria - 02 May 2020; 2020.
\newblock [Online; accessed 11-Aug-2020].
\newblock
  \url{https://www.dhhs.vic.gov.au/coronavirus-update-victoria-02-may-2020}.

\bibitem{streeck2020infection}
Streeck H, Schulte B, Kuemmerer B, Richter E, H{\"o}ller T, Fuhrmann C, et~al.
\newblock Infection fatality rate of SARS-CoV-2 infection in a German community
  with a super-spreading event.
\newblock medrxiv. 2020;.

\bibitem{hamner2020high}
Hamner L.
\newblock High SARS-CoV-2 attack rate following exposure at a choir
  practice—Skagit County, Washington, March 2020.
\newblock MMWR Morbidity and Mortality Weekly Report. 2020;69.

\bibitem{NSW_Aug1}
COVID-19 Weekly Surveillance in NSW, Epidemiological Week 31, Ending 1 August
  2020; 2020.
\newblock [Online; accessed 11-Aug-2020].
\newblock
  \url{https://www.health.nsw.gov.au/Infectious/covid-19/Documents/covid-19-surveillance-report-20200801.pdf}.

\bibitem{SMH_July16}
Aubusson K, Visentin L. Fears of further spread as Crossroads Hotel virus cases
  become infectious within a day; 2020.
\newblock [Online; accessed 11-Aug-2020].
\newblock
  \url{https://www.smh.com.au/national/nsw/fears-of-further-spread-as-sydney-s-crossroads-coronavirus-cases-become-infectious-within-a-day-20200715-p55cds.html}.

\bibitem{NSW_cases}
NSW COVID-19 cases by location and likely source of infection; 2020.
\newblock [Online; accessed 11-Aug-2020].
\newblock
  \url{https://data.nsw.gov.au/data/dataset/nsw-covid-19-cases-by-location-and-likely-source-of-infection}.

\bibitem{DHHS_updates}
Updates about the outbreak of the coronavirus disease (COVID-19); 2020.
\newblock [Online; accessed 11-Aug-2020].
\newblock \url{https://www.dhhs.vic.gov.au/coronavirus/updates}.

\bibitem{sen2020association}
Sen S, Karaca-Mandic P, Georgiou A.
\newblock Association of Stay-at-Home Orders With COVID-19 Hospitalizations in
  4 States.
\newblock JAMA. 2020;.

\bibitem{fair2019creating}
Fair KM, Zachreson C, Prokopenko M.
\newblock Creating a surrogate commuter network from Australian Bureau of
  Statistics census data.
\newblock Sci Data. 2019;6(1):1--14.

\bibitem{ABS_pop18}
1410.0 Data by Region, 2013-18; 2019.
\newblock [Online; accessed 12-Aug-2020].
\newblock
  \url{https://www.abs.gov.au/AUSSTATS/abs@.nsf/DetailsPage/1410.02013-18?OpenDocument}.

\bibitem{viboud2006synchrony}
Viboud C, Bj{\o}rnstad ON, Smith DL, Simonsen L, Miller MA, Grenfell BT.
\newblock Synchrony, waves, and spatial hierarchies in the spread of influenza.
\newblock Science. 2006;312(5772):447--451.

\bibitem{Germann5935}
Germann TC, Kadau K, Longini IM, Macken CA.
\newblock Mitigation strategies for pandemic influenza in the United States.
\newblock Proc Natl Acad Sci USA. 2006;103(15):5935--5940.
\newblock Available from: \url{http://www.pnas.org/content/103/15/5935}.

\bibitem{zachreson2020interfering}
Zachreson C, Fair KM, Harding N, Prokopenko M.
\newblock Interfering with influenza: nonlinear coupling of reactive and static
  mitigation strategies.
\newblock J R Soc Interface. 2020;17(165):20190728.

\bibitem{moss2019can}
Moss R, Naghizade E, Tomko M, Geard N.
\newblock What can urban mobility data reveal about the spatial distribution of
  infection in a single city?
\newblock BMC Public Health. 2019;19(1):1--16.

\bibitem{cliff2018investigating}
Cliff OM, Harding N, Piraveenan M, Erten EY, Gambhir M, Prokopenko M.
\newblock Investigating spatiotemporal dynamics and synchrony of influenza
  epidemics in Australia: An agent-based modelling approach.
\newblock Simul Model Pract Th. 2018;87:412--431.

\bibitem{ferguson2020report}
Ferguson N, Laydon D, Nedjati~Gilani G, Imai N, Ainslie K, Baguelin M, et~al.
\newblock Report 9: Impact of non-pharmaceutical interventions (NPIs) to reduce
  COVID19 mortality and healthcare demand. 2020;.

\bibitem{chang2020modelling}
Chang SL, Harding N, Zachreson C, Cliff OM, Prokopenko M.
\newblock Modelling transmission and control of the COVID-19 pandemic in
  Australia.
\newblock arXiv preprint arXiv:200310218. 2020;.

\bibitem{BingTiles}
Schwartz J. Bing Maps Tile System; 2018.
\newblock [Online; accessed 12-Aug-2020].
\newblock
  \url{https://docs.microsoft.com/en-us/bingmaps/articles/bing-maps-tile-system}.

\bibitem{ABS_TB}
About TableBuilder; 2020.
\newblock [Online; accessed 12-Aug-2020].
\newblock
  \url{https://www.abs.gov.au/websitedbs/d3310114.nsf/home/about+tablebuilder}.

\end{thebibliography}


\end{document}